\documentstyle[preprint,prb,aps]{revtex}
\begin{document}
\draft
\title{Observation of the Fulde-Ferrell-Larkin-Ovchinnikov state in
the quasi-two-dimensional organic superconductor
$\kappa$-(BEDT-TTF)$_2$Cu(NCS)$_2$.}
\author{J. Singleton$^1$, J.A. Symington$^1$, M.-S.~Nam$^1$, A.~Ardavan$^1$,
M.~Kurmoo$^{1,2}$ and P.~Day$^3$}
\address{$^1$Department of Physics, University
of Oxford, The Clarendon Laboratory, Parks Road, Oxford OX1~3PU, UK}
\address{$^2$IPCMS, 23 rue de Loess, BP20/CR, 67037 Strasbourg, France.}
\address{$^3$The Royal Institution, Albemarle Street, London, UK.}
\date{\today}
\maketitle
\begin{abstract}
Single crystals of the
layered organic
type II superconductor, $\kappa$-(BEDT-TTF)$_{2}$Cu(NCS)$_{2}$,
have been studied in magnetic fields of up to 33~T and
at temperatures between 0.5~K and 11~K
using a compact differential susceptometer.
When the magnetic field lies precisely in the quasi-two-dimensional
planes of the material, there is
strong evidence for a phase transition
from the superconducting mixed state
into a Fulde-Ferrell-Larkin-Ovchinnikov (FFLO) state,
manisfested as a change in the rigidity of the vortex system.
The behaviour of the transition as a function of temperature
is in good agreement with theoretical predictions. 
\end{abstract}
\vspace{20mm}
There has been recent renewed interest in the 
Fulde-Ferrell-Larkin-Ovchinnikov (FFLO)~\cite{FF}  
state in superconductors subjected
to high magnetic fields (for a summary,
see Refs.~\cite{Shimahara,others}
and references therein).
In a metal in a magnetic field, the
normal quasiparticles have separate spin-up
and spin-down Fermi surfaces (FSs)
which are displaced due to the Zeeman energy.
In the FFLO state, attractive
interactions of quasiparticles
with opposite spin on 
opposite sides of the two FSs
leads to the formation of pairs with nonzero total
momentum~\cite{FF}; the phase of the gap function varies spatially
with the total momentum, leading to an inhomogeneous
superconducting state.
Calculations show that in anisotropic superconductors
the FFLO might lead to an enhancement of
the upper critical field $B_{\rm c2}$ to
between 1.5 and 2.5 times the
Pauli paramagnetic limit~\cite{Shimahara,others}.

Impetus was added to this theoretical work by
experimental data from the heavy-fermion compounds
CeRu$_2$~\cite{huxley}, UBe$_{13}$~\cite{thomas}
and UPd$_2$Al$_3$~\cite{gloos}, which were initially
interpreted as indications of a FFLO state.
However, subsequent work cast doubt on such
claims. 
In the case of CeRu$_2$, the feature in the magnetisation
ascribed to the FFLO was shown 
to be due to flux-pinning mechanisms involving disorder~\cite{tenya}.
Norman~\cite{norman} also pointed out that the
suggested phase boundary in UPd$_2$Al$_3$
between the superconducting
state and the FFLO does not follow the expected temperature ($T$)
dependence. 

In this paper we describe studies of the resistance
and magnetic behaviour of single
crystals of the quasi-two-dimensional (Q2D)
organic superconductor $\kappa$-(BEDT-TTF)$_2$Cu(NCS)$_2$
which indicate a
phase transition within the
zero-resistance state.
In contrast to the heavy-fermion data~\cite{huxley,thomas,gloos},
the $T$ dependence of this transition
is similar to the predicted
phase boundaries between the superconducting and FFLO states.
  
To induce a FFLO state one needs a
suppression of interactions involving the orbital 
moment (which otherwise destroy
the superconductivity)~\cite{Shimahara},
a FS shape conducive to nesting
(but not enough to induce density-wave states~\cite{shimprb}), and a 
low impurity scattering rate (clean limit)~\cite{shimprb}. 
The first criterion
may be achieved in $\kappa$-(BEDT-TTF)$_2$Cu(NCS)$_2$ 
in an {\it exactly} 
in-plane magnetic field.
The FS of $\kappa$-(BEDT-TTF)$_2$Cu(NCS)$_2$
consists of a warped Q2D cylinder and 
two quasi-one-dimensional sheets~\cite{fss,amro1};
in an in-plane field virtually all the quasiparticle 
paths on the FS are
open orbits~\cite{Shimahara,others,amro1,amro2,msn}. 
Furthermore, it is known that the 
FS of $\kappa$-(BEDT-TTF)$_2$Cu(NCS)$_2$
is prone to nesting~\cite{schmalian};
however, the nesting is insufficient to
cause density-wave states, as shown by
the measured FS 
topology~\cite{fss,amro1}.
Finally, the scattering rates 
and band parameters of $\kappa$-(BEDT-TTF)$_2$Cu(NCS)$_2$ 
suggest that it is in the clean limit~\cite{naka}. 

Single crystals
($\sim 1 \times 0.5 \times 0.1$~mm$^3$; mosaic spread
$\leq \sim 0.1^{\circ}$)
of $\kappa$-(BEDT-TTF)$_{2}$Cu(NCS)$_{2}$
were produced using electrocrystallization~\cite{msn}.
Electrical contacts (resistance $\leq 10~\Omega$)
were made to
the two large faces of each crystal
(parallel to the {\bf bc} (Q2D)
planes)
by attaching $25~\mu$m Au wires or $12.5~\mu$m
Pt wires using graphite paint.
The resistance was measured by driving
an AC current ($5-25~\mu$A, $17-200$~Hz) 
between contacts on the upper
and lower surfaces; the voltage
was measured on an adjacent pair of
contacts using a lock-in amplifier.
In this configuration, the resistance
is proportional to the
interplane resistivity component $\rho_{zz}$~\cite{contacts}.
Care was taken to ensure that 
the measured resistance was neither
frequency nor current dependent.
Individual crystals were mounted in or on
the coil of a tuned-circuit
differential susceptometer (TCDS)~\cite{tcds};
we shall return to the function of the coil below.
The coil was mounted in a cryostat which 
allowed it (and the sample)
to be rotated to all possible orientations
in the magnetic field {\bf B}~\cite{msn}.
The orientation of the sample
is defined by
the polar angle $\theta$ between
{\bf B} and the normal to the
sample's {\bf bc} planes and the azimuthal
angle $\phi$ ($\phi=0$ is
a plane of rotation of {\bf B}
containing {\bf b} and the
normal to the {\bf bc} plane).
The cryostat was placed in 31 and 33~T
Bitter coils at NHMFL, Tallahassee.

Fig.~1(a) shows the superconducting to normal
transition measured in the resistance of a
crystal of $\kappa$-(BEDT-TTF)$_{2}$Cu(NCS)$_{2}$
at $T=4.22$~K and $\theta=90.0^{\circ}$.
Note that the change from zero resistance
to normal-state magnetoresistance occurs over
a range of about 5~T.
This broadened transition region appears to be an intrinsic
feature of (BEDT-TTF)-based superconductors~\cite{review};
it has been attributed to dissipation due to
superconducting weak links in inhomogeneous samples~\cite{itohump,ishihump},
magnetoresistance due to a lattice distortion via
coupling with the quantised vortices~\cite{zuo2} and dissipation caused by
fluctuations characteristic of a $d$-wave superconductor~\cite{maki}.
All of the models mentioned regard the broadened transition
region as an artefact of the superconducting
state; in common with others~\cite{msn,review},
we therefore choose a field $B_{\rm p}$ at which
the extrapolations of the normal-state magnetoresistance
and the transition region intercept (Fig.~1(a))
as a measure of the upper
critical field, $B_{\rm c2}$.

The detection of the FFLO state depends upon the
very precise orientation of the crystal within
the magnetic field, and Fig.~1(b), which displays
the resistance of a
$\kappa$-(BEDT-TTF)$_{2}$Cu(NCS)$_{2}$ sample at $B=26.5$~T
and $T=1.45~$K
as a function of $\theta$, shows how this is accomplished.
In addition to the angle-dependent magnetoresistance
oscillations,
which allow one to determine $\phi$~\cite{fss},
there is a very sharp dip at
$\theta = \pm$~90$^{\circ}$.
The dip occurs because the $\theta$-dependence of the
upper critical field is very sharply peaked at
$\theta=90^{\circ}$~\cite{msn,naka}; $B_{\rm c2}$ is less
than 26.5~T except at angles very close to $\theta=90^{\circ}$.
At exactly $90^{\circ}$, 
$B_{\rm p}$ approaches $30.5$~T at $T=1.45$~K, but
the wide superconducting to normal
transition (see Fig.~1(a)) means that zero resistance is
not attained in Fig.~1(b).
This effect was used to orientate the
crystal; in a fixed applied field which is less
than the maximum possible upper critical field,
but greater than the maximum possible zero-resistance field,
higher values of
$B_{\rm c2}$ lead to lower measured resistances
and {\it vice-versa}.

In order to get the true
in-plane critical field, we found 
it essential to align
the crystal to better than $\pm 0.1^{\circ}$
using the sharp
dip at $\theta=90^{\circ}$;
misalignment by a fraction of a degree
lowers the measured critical field by $\sim 1-2$~T.
We have also found it necessary to work with
{\it small} crystals, as the magnetic forces have
proved capable of bending larger platelets
at orientations close to $\theta=90^{\circ}$,
resulting in the broadening of the
superconducting-normal transition.
In a previous work~\cite{msn}, we showed that the
critical fields of $\kappa$-(BEDT-TTF)$_2$Cu(NCS)$_2$,
whilst varying strongly with $\theta$,
are almost completely independent of $\phi$.

Having orientated the sample, one must
detect the transition between the mixed state
and the FFLO. Ideally, one would use a thermodynamic
measurement of a quantity such as the specific
heat, or detect the change in the vortex arrangement
using neutron scattering or magnetic force microscopy~\cite{rigid}.
However, very considerable difficulties
are inherent in using such techniques
in a tiny single-crystal sample
which is orientated to better than $\pm 0.05^{\circ}$
in a magnetic field $\sim 30$~T at temperatures down to 500~mK.
Moreover, conventional magnetometry
is unhelpful in this context, as there is no discontinuous
change in the amount of flux penetrating
the sample at the transition~\cite{Shimahara,rigid}.

We have chosen instead to examine the {\em rigidity}
of the vortex arrangement, which has been predicted
to change on going from mixed state to FFLO~\cite{rigid}.
The sample is mounted with its Q2D planes perpendicular
to the axis of the TCDS coil.
When the quasistatic magnetic field
is in the sample planes ($\theta=90^{\circ}$),
the TCDS coil provides 
a small oscillating magnetic field {\it perpendicular}
to the static field (and the vortices) which
exerts a torque
on the vortices.
The coil in the TCDS forms part of a tank circuit
operating at $\sim 3$~MHz, so that changes in the rigidity
of the vortices ({\it i.e.} their 
resistance to the torque) affect the effective ``stiffness'' of the
circuit and therefore shift its resonant frequency~\cite{tcds}.
 
Fig.~2(a) shows simultaneous measurements of
the TCDS frequency and sample resistance as a function of
magnetic field ($\theta =90.00 \pm 0.05^{\circ}$, $\phi=0$).
The resistive
onset is at $\sim 28$~T ($B_{\rm p} \approx 33.5$~T).
The TCDS frequency has been normalised by dividing by
the field dependence of the frequency of the empty coil
(caused by the magnetoresistance of the copper windings~\cite{tcds});
any remaining field dependence is due to the
presence of the sample.
Superimposed on the gentle downward trend, which
results from the growing flux penetration~\cite{tcds},
is a steeper drop, or ``elbow'', starting at $\sim 23$~T.
The drop occurs below the resistive onset (and well below $B_{\rm p}$);
it therefore represents a change of sample behaviour
{\it within} the superconducting state.
We label the field at which the drop
occurs $B_{\rm L}$,
defining it using the intersection of the
extrapolations shown in Fig.~2(a).
As the temperature rises, $B_{\rm L}$ moves to
lower magnetic fields, and
Fig.~2(b) shows some representative data.
The ``elbow'' in the field dependence of the
TCDS frequency is visible up to about $T\approx 5.4$~K
(arrow in Fig.~2(b)),
but could not be detected at higher temperatures.

The fields $B_{\rm p}$ and $B_{\rm L}$
are shown as a function of $T$
in Fig.~3 for a number of samples.
The $B_{\rm p}$ data are in good agreement with previous
studies of $B_{\rm c2}$ in in-plane fields~\cite{msn,newzuo},
once one makes allowance for the different
methods of defining the upper critical field
employed in the earlier works;
the high $B_{\rm p}$ values obtained
in this work ($B_{\rm p}(T=0) \approx 35$~T)
are indicative of the high quality of our crystals,
their precise orientation and their freedom from strain
or bending.
Note that there were vestiges of the
superconducting transition visible 
at $T \sim 10.4$~K in our samples, although zero
resistance was attained only below $T \sim 9.6$~K.

Before discussing Fig.~3 further, we consider whether the feature at
$B_{\rm L}$ might be explained
by the vortex-related
instabilities which have been invoked to dismiss
claims for the FFLO state in heavy-fermion
materials~\cite{tenya,schimanski}.
Such mechanisms were also probably responsible
for an earlier erroneous claim of a FFLO in
$\kappa$-(BEDT-TTF)$_{2}$Cu(NCS)$_{2}$~\cite{ballsup}.
These instabilities are describable by the
``synchronisation pinning scenario'' (SPS)~\cite{steingart}
and may be identified by their history dependence~\cite{tenya,steingart}
and the way in which the $T$ dependence
of the instability field follows that of $B_{\rm c2}$,
falling to zero at, or just below $T_{\rm c}$ (see also 
Ref.~\cite{norman}).
By contrast, the $T$ dependence of $B_{\rm L}$
is completely different than that of $B_{\rm p}$, and
by inference, $B_{\rm c2}$ (Fig.~3).

Experiments involving a number of different
samples, thermal cycling and the application of
bias voltages to the samples
were carried out to assess
whether the SPS has a role in the feature at
$B_{\rm L}$; data from three samples subjected
to different bias and cooling conditions are shown in Fig.~3.
The measured $B_{\rm L}(T)$ does not
seem to be affected by thermal cycling,
cooling rate, bias current and route into the superconducting state
to within experimental accuracy.
This is in sharp contrast to the behaviour of
vortex-related (SPS) instabilities~\cite{tenya,steingart},
and strongly suggests that the feature at $B_{\rm L}$
is an intrinsic effect in $\kappa$-(BEDT-TTF)$_2$Cu(NCS)$_2$.
By contrast, an earlier feature which had been tentatively
associted with the FFLO~\cite{ballsup} was found to disappear on thermal
cycling. 

A shift in vortex rigidity is one of the
key identifying features of the FFLO state~\cite{schimanski}.
Fig.~3 therefore compares $B_{\rm L}$ and
$B_{\rm p}$ with
the calculated FFLO phase diagram of Ref.~\cite{shimprb},
derived for a generic Q2D metal.
The theoretical curves have been
scaled using a $T=0$ $B_{\rm c2}$ of 35~T
and $T_{\rm c}=10$~K.
Even though there are (not unexpected~\cite{itohump,ishihump,zuo2})
deviations of $B_{\rm p}$ from
the theoretical dependence of $B_{\rm c2}$,
the experimental data in Fig.~3 bear a 
striking similarity to the calculated phase boundaries
of Ref.~\cite{shimprb}.
In particular, $B_{\rm L}$ follows the phase boundary
between the Type-II superconducting state
and the FFLO state (dotted curve) closely, extrapolating
to $B_{\rm p}$
at $T \sim T^*=0.56T_{\rm c}$.
The meeting of the two phase boundaries
at $T^*=0.56_{\rm c}$ is a robust feature
of models of the FFLO,
irrespective of dimensionality~\cite{shimprb}.

The sensitivity of the FFLO state to orbital effects has been
mentioned above.
Fig.~4 shows the TCDS frequency versus magnetic field
for several angles $\theta=90^{\circ}\pm \Delta \theta$.
It is obvious that the ``elbow'' at $B_{\rm L}$,
identified with the presence of the FFLO,
is only visible for
$|\Delta \theta | <\sim 1.5^{\circ}$;
this is in good agreement with the theoretical calculations
of Refs.~\cite{rigid,austrians}, which predict that the LOFF is only
stable in typical organic
conductors for $|\Delta \theta| < \sim 0.3-2.3^{\circ}$.
For bigger deviations than this, a substantial
number of closed orbits will be possible
on the FS~\cite{amro1,amro2}, leading to suppression
of the superconductivity due to orbital effects~\cite{msn,rigid,newzuo,austrians}.
It is notable that several groups have found it difficult
to fit the angle-dependence of $B_{\rm c2}$
close to $\theta=90^{\circ}$ (see Ref.~\cite{review}
for a discussion and comparison of the various formulae employed);
the presence of the FFLO and the resulting increase in
upper critical field may well account for this.
Note that the number of closed orbits at a particular
value of $\theta$ will depend on $\phi$~\cite{amro1,amro2},
so that the range of $\theta$ over which the FFLO can exist
may depend critically on $\phi$.
Only a restricted number
of values of $\phi$ have been examined in detail thus far,
and it is difficult to make more than a tentative judgement,
but we remark that the region of $\theta$ over which
the feature at $B_{\rm L}$ is visible appears
to become narrower as $\phi$ increases from
$0$ towards $90^{\circ}$.

In summary, we have measured
the magnetic behaviour and resistivity
of single crystals of the organic
superconductor, $\kappa$-(BEDT-TTF)$_{2}$Cu(NCS)$_{2}$.
When the field lies in the Q2D
planes of the material, there is
evidence for a phase transition
into what we believe is a Fulde-Ferrell-Larkin-Ovchinnikov (FFLO) state.
The phase transition is in good agreement with theoretical predictions.

This work is supported by EPSRC (UK).
NHMFL is supported by the
Department of Energy, the National
Science Foundation and the State of Florida.

\clearpage

\begin{figure}
\vspace{150mm}
\includegraphics{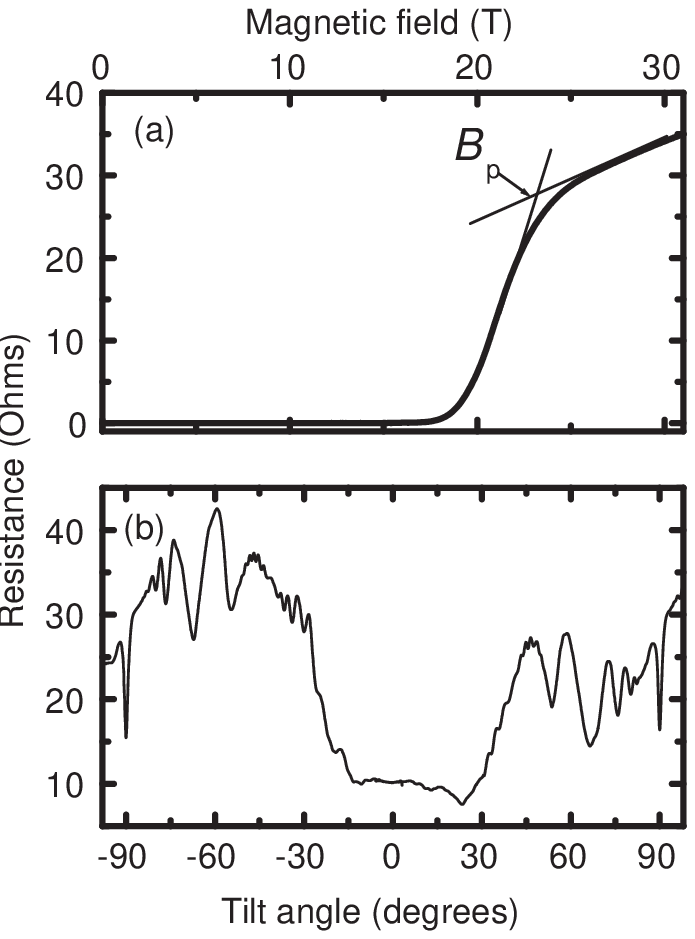}
\caption{(a) Resistance of a single
crystal of $\kappa$-(BEDT-TTF)$_2$Cu(NCS)$_2$
as a function of magnetic field
at $T=4.22$~K and $\theta=90.0^{\circ}$.
Note the broadened transition from zero resistance to
normal-state magnetoresistance.
The intersection of the straight lines defines
the critical field $B_{\rm p}$.
(b)~Resistance of a $\kappa$-(BEDT-TTF)$_2$Cu(NCS)$_2$
crystal at $\phi = 150^{\circ}$, $B=26.5$~T and $T=1.45$~K,
as a function of tilt angle $\theta$.
The data show behaviour typical of the
normal-state magnetoresistance
({\it e.g.} angle-dependent magnetoresistance oscillations
and an asymmetric
rise in resistance as $\theta$ increases from
0 towards $90^{\circ}$)~\cite{amro1,amro2}
except very close to $\theta=90^{\circ}$, where the sharp
dips indicate the onset of superconductivity. 
}
\end{figure}

\clearpage
~

\begin{figure}
\vspace{150mm}
\includegraphics{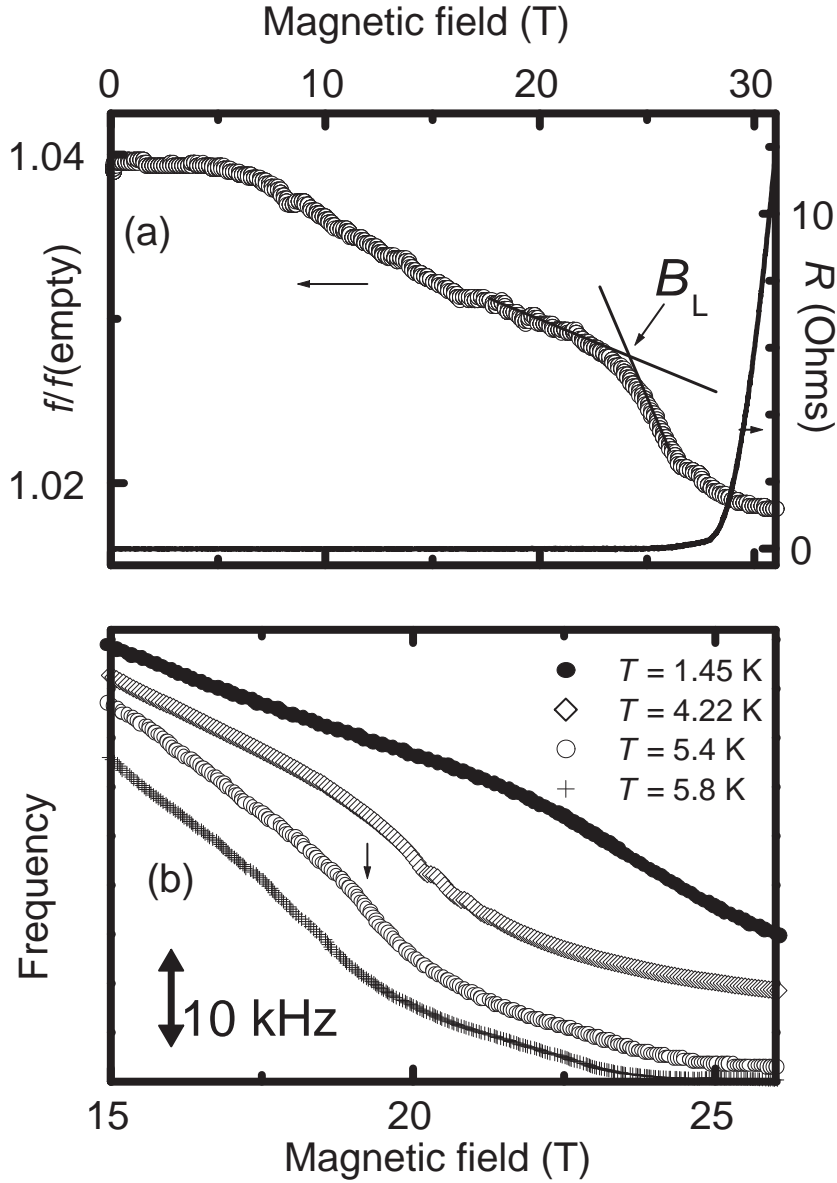}
\caption{(a) Normalised TCDS frequency ({\it i.e.}
frequency of loaded coil divided by frequency
of the empty coil) (points, left axis)
and sample resistance (solid curve, right axis) versus
magnetic field; $T=0.44$~K, $\theta=90^{\circ}$,
$\phi=0^{\circ}$.
The intersection of the two lines at the ``elbow''
in the TCDS response
defines the field $B_{\rm L}$.
(b) Raw TCDS frequency versus magnetic field
for temperatures $T =1.45$, 4.22, 5.4 and 5.8~K,
showing the movement of $B_{\rm L}$ to higher
fields as $T$ decreases ($\phi=0^{\circ}$);
the data have been vertically offset for clarity.
The arrow indicates the ``elbow'' in the frequency
at $B_{\rm L} \approx 19$~T for $T=5.4$~K; for higher
temperatures ({\it e.g.} 5.8~K) it was not
observed.
}
\end{figure}

\clearpage
~

\begin{figure}
\vspace{150mm}
\includegraphics{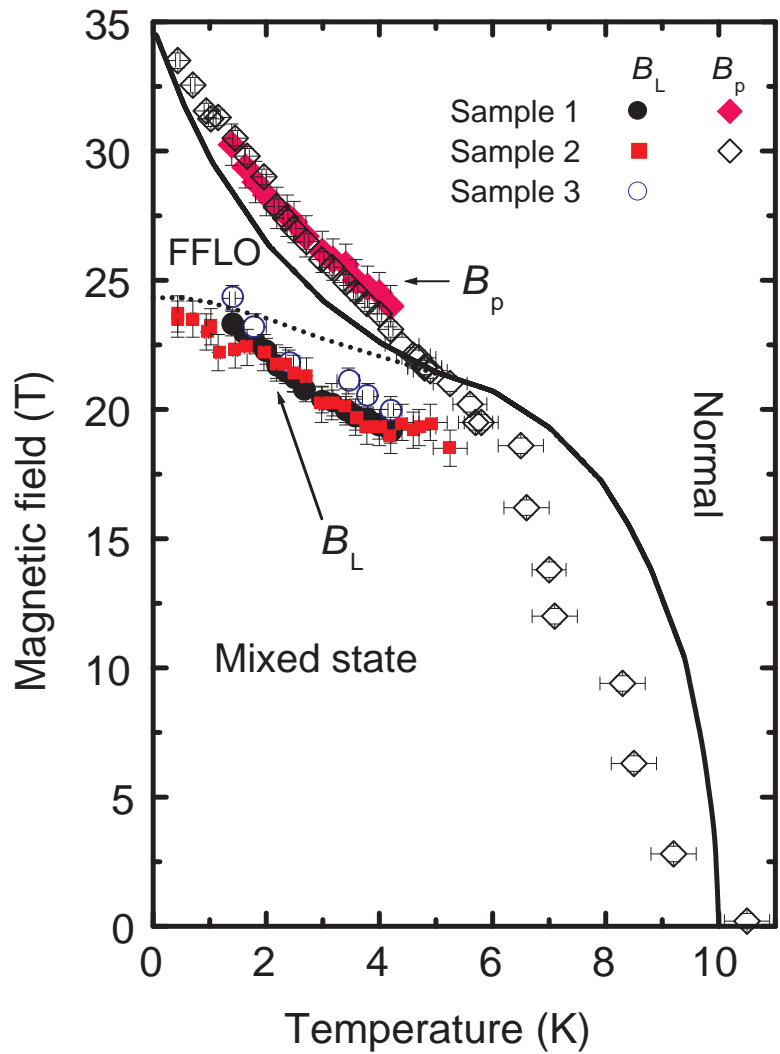}
\caption{Temperature dependence of
the fields $B_{\rm L}$ and 
$B_{\rm p}$ at
$\theta=90^{\circ}$, $\phi=0^{\circ}$, compared with
the FFLO phase diagram of Ref.~[10];
the solid curve separates the
superconducting and normal states.
The boundary between the mixed state and
the FFLO is shown as a dotted line.
Data for three samples under differing conditions
of bias and thermal cycling are shown.}
\end{figure}

\clearpage
~

\begin{figure}
\vspace{150mm}
\includegraphics{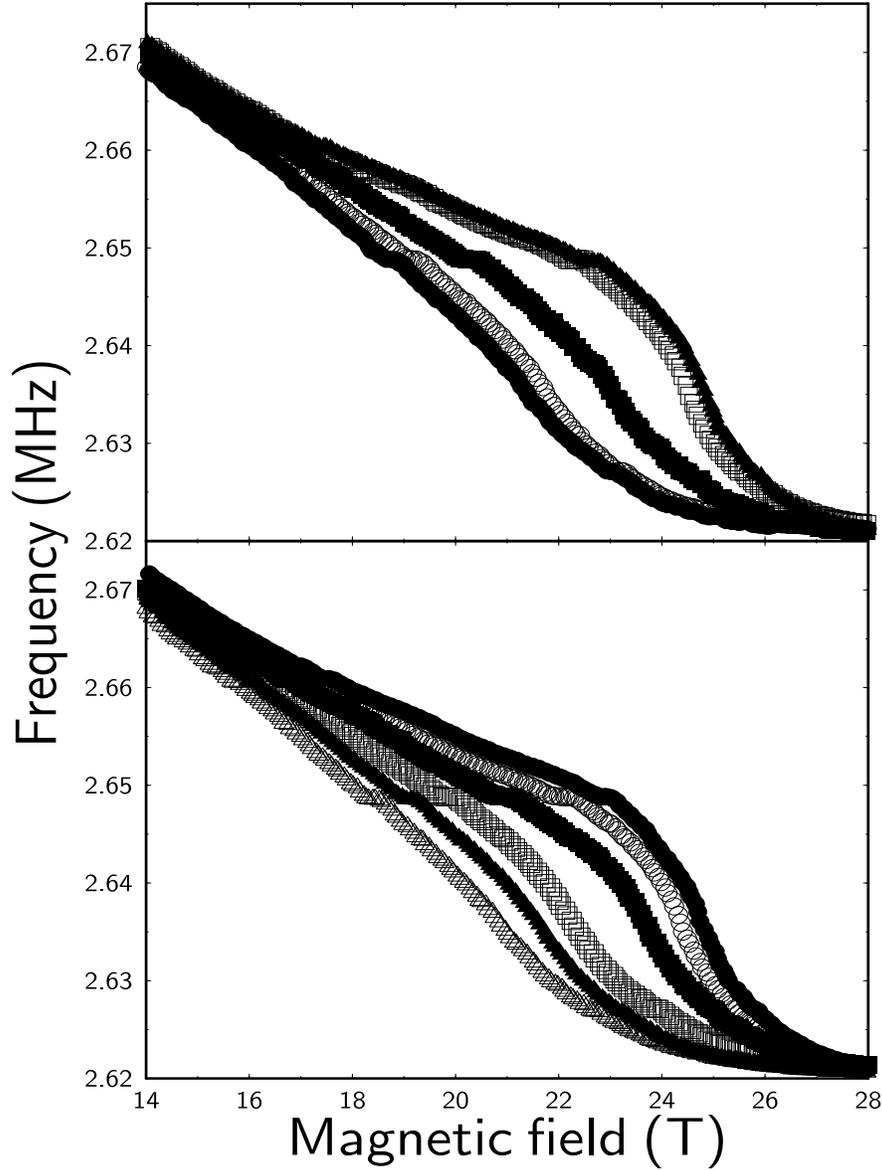}
\caption{Raw TCDS frequency versus magnetic field
for several different values of $\theta$. Upper:
black triangles,~90.15$^\circ$; 
open squares~89.74$^\circ$;
black squares~89.17$^\circ$; 
open circles~88.61$^\circ$;
black circles~88.33$^\circ$. 
Lower: 
black circles~90.00$^\circ$;
open circles~90.35$^\circ$;
black squares~90.74$^\circ$; 
open squares~91.12$^\circ$;
black triangles,~91.53$^\circ$; 
open triangles,~91.92$^\circ$.
$\phi$ is -45$^{\circ}$.
The elbow at $B_{\rm L}$ disappears when the
angle differs from $90^{\circ}$ by more than
about $\sim 1.5^{\circ}$.
Note that the small step seen in all data just below
2.65~MHz is an artefact of an internal range-change in the
electronic frequency meter; it is present under all
experimental conditions at the same frequency.}

\end{figure}

\end{document}